\documentclass[conference,letterpaper]{IEEEtran}

\usepackage[utf8]{inputenc} 
\usepackage[T1]{fontenc}
\usepackage{url}
\usepackage{ifthen}
\usepackage{cite}
\usepackage[cmex10]{amsmath} 
\usepackage[dvipdfmx]{graphicx}
\usepackage{latexsym}
\usepackage{amsthm}
\usepackage{ascmac}
\usepackage{xcolor}
\usepackage{enumitem}
\newtheorem{theorem}{Theorem}
\newtheorem{definition}{Definition}
\newtheorem{remark}{Remark}
\newtheorem{lemma}{Lemma}

\renewcommand{\figurename}{Fig.}
\renewcommand{\refname}{References}

\newcommand{\defeq}{\mathrel{\mathop:}=}

\begin{document}
\title{Identification, Secrecy, Template, and Privacy-Leakage of Biometric Identification System Under Noisy Enrollment}


\author{%
  \IEEEauthorblockN{Vamoua Yachongka}
  \IEEEauthorblockA{Dept. of Computer and Network Engineering\\
                    The University of Electro-Communications\\
                    Tokyo, Japan\\
                    Email: va.yachongka@uec.ac.jp}
  \and
  \IEEEauthorblockN{Hideki Yagi}
  \IEEEauthorblockA{Dept. of Computer and Network Engineering\\
                    The University of Electro-Communications\\ 
                    Tokyo, Japan\\
                    Email: h.yagi@uec.ac.jp}
}


\maketitle

\begin{abstract}
In this study, we investigate fundamental trade-off among identification, secrecy,
template, and privacy-leakage rates in biometric identification systems. Ignatenko
and Willems (2015) studied this system assuming that the channel in the enrollment process
of the system is noiseless and they did not consider the template rate. In the enrollment process, however,
it is highly considered that noise occurs when bio-data is scanned. In this paper, we impose
a noisy channel in the enrollment process and 
characterize the capacity region of the rate tuples. The capacity region is proved by
a novel technique via two auxiliary random variables, which has never been seen in previous studies.
As special cases, the obtained result shows that
the characterization reduces to the one given by Ignatenko and Willems (2015)
where the enrollment channel is noiseless and there is no constraint on the template rate, and
it also coincides with the result derived by G\"unl\"u and Kramer (2018) where
there is only one individual.
\end{abstract}



\section{Introduction}

Biometric security is a security mechanism used to identify an individual on
the basis of his/her physical characteristics.
Biometric technology enables us to recognize the individual by matching the unique feature with
biological data (bio-data) already stored in the system database.
Some well-known technologies of this kind of security are fingerprint-based
identification, iris-based identification, voice recognition, etc.
Nowadays many applications make use of
this technology like homeland checking at land port, mobile payment with smartphone and so on.

O'Sullivan and Schmid \cite{OS} and Willems et al.\ \cite{willems} independently introduced the
discrete memoryless {\em biometric identification system} ({BIS}). Basically, the BIS consists of two
phases: (I) {\em Enrollment Phase} and (I\hspace{-.1em}I)
{\em Identification Phase}. In (I) Enrollment Phase, all individuals' bio-data sequences are
generated from a stationary memoryless source. The sequences are observed through a noisy
{\em discrete memoryless channel} ({DMC})
and stored into system database.
In (II) Identification Phase, a bio-data sequence of an
unknown individual is observed via another noisy DMC, and
an estimated value of the unknown individual is output.

There are many studies related to the BIS. We highlight some previous studies which
are {deeply} connected to this study.
Willems et al.\ \cite{willems} have clarified
the identification capacity of the BIS, which is the maximum achievable rate of the number of
individuals when the error probability converges to zero as the length of {bio-data}
sequences goes to infinity. However, the {system model} in \cite{willems}
{assumes that}
bio-data sequences are stored in the system database in a plain form, leading
to a critical privacy leakage threat.
Tuncel \cite{tuncel} has extended their model by
incorporating compression of bio-data sequences stored in the system database and
clarified the capacity region of identification and coding rates
({in} this study, a codeword is called a template, and this coding rate is called the template rate).
Later, Ignatenko and Willems
\cite{itw3} investigated the BIS model with secret data and template generation.
Related to this work, the system with only secrecy estimation has been {analyzed} in
\cite{itwl1}--\hspace{-0.1mm}\cite{koide}. In \cite{itw3},
the authors evaluated the amount
of information leaked between a template stored in {the} database and its bio-data sequence,
called the privacy-leakage rate, and
clarified the fundamental trade-off among identification, secrecy and privacy-leakage
rates in the BIS provided that the enrollment channel is {\em noiseless}.
Recently, Yachongka and Yagi \cite{vy}
introduced a constraint of the template rate to the model developed in \cite{itw3} and
clarified the fundamental trade-off among identification, {secrecy, and} template rates in the BIS.

An interesting observation {given in \cite{itw3} for the case where the secrecy rate is zero
and in \cite{vy} for the case where the secrecy rate is} positive {indicates} that the {minimum}
{required amount of the} template
rate is equal to the minimum {required amount of the} privacy-leakage rate when
the enrollment channel is {{\em noiseless}}.
Despite this insight, when bio-data is scanned and stored
in the system database, it is highly considerable that bio-data sequences
are subject to noise, as is assumed in \cite{willems}, \cite{tuncel}, and \cite{vy1}.
Actually, by treating {a} {\em noisy} {enrollment channel},
{the problem becomes} more challenging and interesting,
especially, in the evaluation of the privacy-leakage rate.
This motivates us to consider a noisy channel in the enrollment phase of {the} BIS.

In this paper, we aim to {characterize} the capacity region of identification, secrecy, template, and privacy-leakage rates
in the BIS. In order to get closer to practical system, we analyze the region by imposing
the following requirements:
\begin{enumerate}[label=\arabic*)]
	\item there is a {{\em noisy}} channel in the enrollment phase,
	\item we consider a scheme of both protecting privacy (as in \cite{itw3})
        and compressing template (as in \cite{tuncel} and \cite{vy}),
	\item we analyze the capacity region provided that the prior distribution of an identified individual is unknown.
\end{enumerate}

To handle the difficulties of bounding the privacy-leakage rate {in the achievability proof},
we introduce a {\em virtual} system with a {\em partial} decoder,
which outputs only the secret data of individual.
We show that there are two different ways to express the capacity region of the BIS.
An expression uses a single auxiliary random variable (RV) and another requires
two auxiliary RVs.
Later, we will {demonstrate} that the two regions (regions with one and two auxiliary RVs)
are technically identical in Remark \ref{remark2222}.
Although there are two different aspects,
we provide the proof of our main result based on the one employing two auxiliary RVs.
Some benefits of deriving via two auxiliary RVs are {that} the achievability proof can be done
{in} a {simpler} form since each rate constraint is addressed individually.
The characterization of the capacity region of the BIS is basically similar to the ones
given in \cite[Theorem 1]{itw3}, \cite[Theorem 1]{onur}, and \cite{vy}.
As special cases, it can be checked that our characterization reduces to the one given
by Ignatenko and Willems \cite[Theorem 1]{itw3}
where the enrollment channel is noiseless and there is no constraint on the template rate, and
it also coincides with the result derived by G\"unl\"u and Kramer \cite[Theorem 1]{onur} where there is
{only one individual, and thus individual's estimation is not necessary}.

The rest of this paper is organized as {follows}. In Sect.\ \ref{sec2},
we define notation used in this paper and describe the details of the system model.
In Sect.\ \ref{sec3}, {we present} our main result. Next,
we provide the {detailed} proofs of the main result in Sect.\ \ref{sec4}. Finally,
in Sect.\ \ref{sec5}, we give some concluding remarks and future works.

\section{System Model} \label{sec2}
In this section, we define notation used in this paper and describe the details
of the system model {within} information theoretic framework.
\subsection{Notation}
Calligraphic $\mathcal{A}$ stands for a finite alphabet. Upper-case $A$ denotes a RV
taking values in
$\mathcal{A}$ and lower-case $a \in \mathcal{A}$ denotes its realization.
$P_A(a)~\defeq~\Pr[A = a]$, $a \in \mathcal{A}$, represents 
{the} probability distribution on $\mathcal{A}$, and $P_{A^n}$ represents {the} probability distribution 
of RV $A^n = (A_1,\cdots,A_n)$ in $\mathcal{A}^n$, the {$n$th} Cartesian product of 
$\mathcal{A}$. $P_{A^nB^n}$ represents the joint probability distribution of a pair of
RVs $(A^n,B^n)$ and its conditional probability distribution
$P_{A^n|B^n}$ is defined as
\begin{align}
&P_{A^n|B^n}(a^n|b^n) = \frac{P_{A^nB^n}(a^n,b^n)}{P_{B^n}(b^n)}\nonumber \\
&~~~~(\forall a^n \in \mathcal{A}^n, \forall b^n \in \mathcal{B}^n ~\mathrm{such}~\mathrm{that}~P_{B^n}(b^n)~>~0).
\end{align}

The entropy of RV $A$ is denoted by $H(A)$, the joint entropy of RVs $A$ and $B$ 
is denoted by
$H(A,B)$, and the mutual information between $A$ and $B$ is denoted by $I(A;B)$ \cite{cover}.
Throughout this paper, logarithms are of base two.
For integers $a$ and $b$ such that
$a \leq b$, $[a,b]$ denotes the set 
$\{a,a+1,\cdots,b\}$. A partial sequence of a sequence $c^n$ from the first symbol
to the $t$th symbol $(c_1,\cdots,c_{t})$ is represented by $c^{t}$.

Here, we define the strong typicality property and use the same notation as in \cite{cover}.
A sequence $x^n \in \mathcal{X}^n$ is said to be $\delta$-$strongly~typical$ with
respect to a distribution {$P_{X}$} on $\mathcal{X}$ if $|\frac{1}{n}N(a|x^n) - P_{X}(a)| \leq \delta$ and
$P_{X}(a) = 0$ implies $\frac{1}{n}N(a|x^n) = 0$ for all $a \in \mathcal{X}$, where $N(a|x^n)$
is the number of occurrences of $a$ in the sequence $x^n$, and $\delta$ is
an arbitrary positive number.
The set of sequences $x^n \in \mathcal{X}^n$ such that $x^n$ is $\delta$-strongly typical is called
the strongly typical set and is denoted by $A^{(n)}_{\epsilon}(X)$. This concept is easily extended to
joint distributions.

\subsection{Model Descriptions}
\begin{figure}[!h]
 \begin{center}
  \includegraphics[width = 85mm]{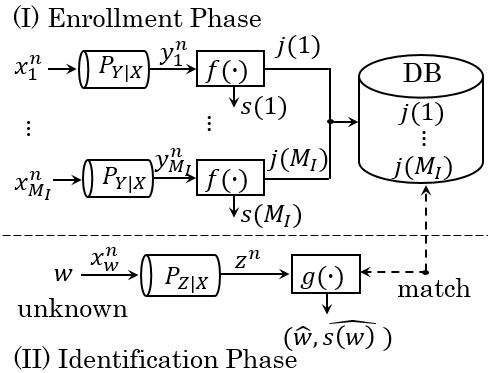}
 \end{center}
 \caption{BIS model}
 \label{fig:gdmc}
\end{figure}

The BIS model studied in this paper is shown in Fig.\ \ref{fig:gdmc}. It
consists of two phases: (I) {\em Enrollment Phase}, and
(I\hspace{-.1em}I) {\em Identification Phase}. Next, {we} will explain the details of each phase. 
\vspace{-3mm}

\noindent{(}I) {\em Enrollment Phase}:

Let $\mathcal{I} = [1,M_I]$ and $\mathcal{X}$ be the sets of indexes of individuals and
a finite source alphabet, respectively.
For any $i \in \mathcal{I}$, we assume that
$x^n_i = (x_{i1},\cdots ,x_{in}) \in \mathcal{X}^n$, an $n$-length
bio-data sequence of individual $i$, is generated 
i.i.d. from a stationary memoryless source $P_X$. The generating probability for each 
sequence $x^n_i \in \mathcal{X}^n$ is
\begin{align}
P_{{X^n_i}}(x^n_i)~\defeq~\Pr[{X^n_i} = x^n_i]~=~\displaystyle \prod_{\substack{k = 1}}^nP_{X}(x_{ik}).
\label{dataseisei}
\end{align}

Now let $\mathcal{J} = [1,M_J]$ and $\mathcal{S} = [1,M_S]$
be the {sets} of indexes of templates stored in database and individuals' secret data, respectively.
All bio-data sequences are observed via a stationary DMC
$\{\mathcal{Y},P_{Y|X},\mathcal{X}\}$, where
$\mathcal{Y}$ is a finite output-alphabet of $P_{Y|X}$.
The corresponding probability that $x^n_i \in \mathcal{X}^n$
is observed as $y^n_i = (y_{i1},y_{i2},\cdots ,y_{in}) \in \mathcal{Y}^n$ via the DMC $P_{Y|X}$ is
\begin{align}
{P_{Y^n_i|X^n_i}(y^n_i|x^n_i)}~=~\displaystyle \prod_{\substack{k = 1}}^nP_{Y|X}(y_{ik}|x_{ik}) 
\label{eq2}
\end{align}

\noindent{f}or all $i \in \mathcal{I}$. Afterwards, the observed bio-data sequence $Y^n_i$
is encoded into template $J(i) \in \mathcal{J}$ and secret data $S(i) \in \mathcal{S}$ as
\begin{align}
(J(i),S(i)) = f(Y^n_i)~~~~(i \in \mathcal{I}),
\end{align}
where
$
f:~\mathcal{Y}^n \longrightarrow \mathcal{J} \times \mathcal{S}
$
denotes {encoding function}. The corresponding template {$J(i)$} is a compressed version of sequence $Y^n_i$ and
stored at position $i$ in {the} database, which can be accessed
by the decoder. Contrarily, the secret data {$s(i)$} is returned to individual $i$ and kept
as confidential. We denote the database as $\mathcal{J}_{M_I}=\{J(1),\cdots,J(M_I)\}$ for brevity purpose
in the upcoming analyses.

\medskip
\noindent{(I\hspace{-.1em}I) {\em Identification Phase}}:

Bio-data sequence $x_w^n\ (w \in\mathcal{I})$ of an unknown $w$ ({index of individual enrolled} in
{the} database)
is observed via a DMC $\{\mathcal{Z},P_{Z|X},\mathcal{X}\}$, where $\mathcal{Z}$ is a finite output-alphabet of 
$P_{Z|X}$. The corresponding probability that $x^n_w \in \mathcal{X}^n$ is output as $z^n = (z_{1},z_{2},\cdots ,z_{n}) \in \mathcal{Z}^n$
via $P_{Z|X}$ is given by
\begin{align}
{P_{Z^n|X^n_w}(z^n|x^n_w)}~=~\displaystyle \prod_{\substack{k = 1}}^nP_{Z|X}(z_k|x_{wk}). \label{eq3}
\end{align}
The decoder observers the identified sequence $Z^n$ and estimates the pair of
index and secret data by comparing $Z^n$ with all templates $\mathcal{J}_{M_I}$ in the database
$(\widehat{W},\widehat{S(W)}) = g(Z^n,\mathcal{J}_{M_I})$, where $g$ denotes decoding function.

\begin{remark}
Note that the distribution of $P_{X}$, $P_{Y|X}$, and {$P_{Z|X}$} are assumed to be known or fixed
and RV $W$ is independent of $(X^n_i,Y^n_i,J(i),S(i),Z^n)$ for all $i \in \mathcal{I}$ like
previous studies. But, in this paper we assume neither that the identified individual
index $W$ are uniformly distributed over $\mathcal{I}$ nor that there is a prior distribution of
{$W$}.
\end{remark}

The motivation {to analyze} performance of the BIS provided that the distribution of
$I$ is unknown is that the identified frequencies of each individual
are likely different. For example, it is hard to think that the frequencies of coming
to use a bank teller of each {individual} are identical. {For} real {applications},
this assumption is important to take care of.

\section{Definitions and Main Results} \label{sec3}

The formal definition and main theorem of this study are given below.
\begin{definition} \label{def1}
The tuple of an identification, secrecy,
template, privacy-leakage rates $(R_I,R_S,R_J,R_L)$ is said to be achievable if for any $\delta > 0$
and large enough $n$ there exist pairs of encoders and decoders that {satisfy}
\begin{align}
\textstyle \max_{\substack{i \in \mathcal{I}}} \Pr\{{(\widehat{W},\widehat{S(W)})} &\neq (W,S(W))|W=i\} \leq  \delta, \label{a} \\
\textstyle \frac{1}{n}\log{M_I} &\geq   R_I - \delta{,} \label{b} \\
\textstyle \min_{\substack{i \in \mathcal{I}}} \frac{1}{n}H(S(i)) &\geq R_S - \delta, \label{c} \\
\textstyle \frac{1}{n}\log{M_J} &\leq   R_J + \delta, \label{d} \\
\textstyle \max_{\substack{i \in \mathcal{I}}} \frac{1}{n}I(S(i);J(i)) &\leq  \delta, \label{shiki4} \\
\textstyle \max_{\substack{i \in \mathcal{I}}} \frac{1}{n}I(X^n_i;J(i)) &\leq   R_L + \delta. \label{dd}
\end{align}
Moreover, the capacity region $\mathcal{R}$ is defined as the closure of the set of all achievable
rate tuples.
\end{definition}
In Definition \ref{def1}, (\ref{a}) is the condition of the error probability
of an individual $i$, which is arbitrarily small.
Equations {(\ref{b})--(\ref{d})} are
the constraints related to identification, {secrecy, and template} rates, respectively.
In term of the privacy protection perspective, we measure the information leakage of individual $i$ by
{(\ref{shiki4}) and (\ref{dd})}.
Condition (\ref{shiki4}) measures the {secrecy-leakage}
between the template in the database and the secret data of individual $i$, and
it requires that {the maximum leaked amount} is not greater than $\delta$.
Condition (\ref{dd}) measures the amount of {privacy-leakage} of original bio-data $X^n_i$
from template {$J(i)$} and its maximum value must be smaller than or equal to {$R_L + \delta$}.
\begin{remark}
In \cite{itw3}, a {stronger} requirement that the distribution of secret data
of every individual must be almost uniform, i.e.
${\frac{1}{n}H(S(i))} + \delta \ge \frac{1}{n}\log M_S$, is included in (\ref{c}).
However, {this} requirement was not
actually necessary in the general problem {formulation}.
\end{remark}


\begin{theorem}\label{th1}
The capacity region for the BIS is given by
\vspace{-2mm}
\begin{align}
\mathcal{R} = \mathcal{A}_1,
\end{align}
\vspace{-2mm}
\noindent{w}here $\mathcal{A}_1$ is defined as
\begin{align}
\hspace{-8mm} \mathcal{A}_1 = \bigcup_{P_{U|X}}\{(R_I,&R_S,R_J,R_L):~R_I + R_S \leq I(Z;U),\nonumber \\
&R_J \geq I(Y;U) - I(Z;U) + R_I,\nonumber \\
&R_L \geq I(X;U) - I(Z;U) + R_I, \nonumber \\
&R_I \geq 0, R_S \geq 0\}, \label{theorem1}
\end{align}
where auxiliary RV $U$ takes values in a
finite alphabet $\mathcal{U}$ with $|\mathcal{U}| \leq |\mathcal{Y}| + 2$.
\qed
\end{theorem}

\begin{remark} \label{remark2222}
We define a region $\mathcal{A}_2$ as
\begin{align}
\hspace{-6mm} \mathcal{A}_2 = \bigcup_{P_{U|X},P_{V|U}}\{(R_I,&R_S,R_J,R_L):\nonumber \\
0 \leq~& R_I \leq I(Z;V),\nonumber \\
0 \leq~& R_S \leq I(Z;U)-I(Z;V),\nonumber \\
&R_J \geq I(Y;U) - I(Z;U) + I(Z;V),\nonumber \\
&R_L \geq I(X;U) - I(Z;U) + I(Z;V)\}, \label{theorem2}
\end{align}
\noindent{w}here auxiliary RVs $U$ and $V$ take values in some
finite alphabets $\mathcal{U}$ and $\mathcal{V}$
with $|\mathcal{U}| \leq (|\mathcal{Y}| + 2)(|\mathcal{Y}| + 3)$ and $|\mathcal{V}| \leq |\mathcal{Y}|+ 3$.
Then, it can be verified that
\begin{align}
\mathcal{A}_1 = \mathcal{A}_2 \label{region333}
\end{align}
for which the proof is given in Appendix A. In this {paper}, we will
{prove} Theorem \ref{th1} based on
the {rate constraints} of the region $\mathcal{A}_2$ instead of $\mathcal{A}_1$.
\end{remark}

\begin{figure}[h]
   \vspace{-3mm}
\begin{center}
   \includegraphics[width=85mm]{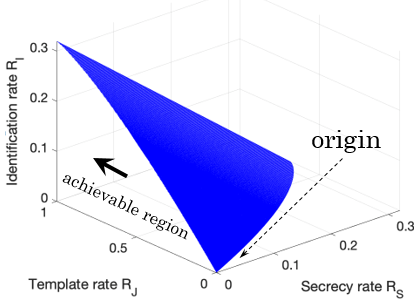}
   \caption{The rate region of the BIS}
    \label{tcp1}
 \end{center}
\begin{center}
   \includegraphics[width=85mm]{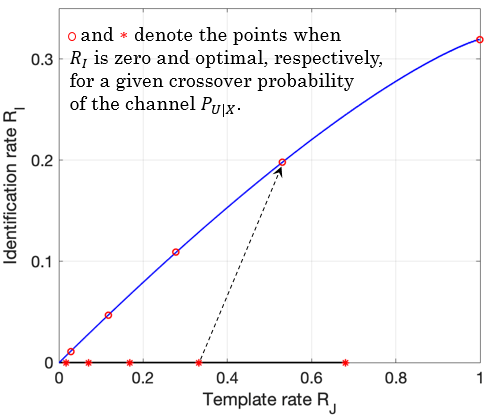}
   \caption{Projection of the rate region onto $R_JR_I$-{plane}}
   \vspace{-3mm}
    \label{tcp2}
 \end{center}
\end{figure}

As we have previously mentioned, one can check that the characterization of Theorem 1 coincides with
the region characterized by Ignatenko and Willems \cite[Theorem 1]{itw3} {in two steps:
first replace $Y$ by $X$ and then remove the constraint $R_J$ from (\ref{theorem1}).
The obtained region {is identical} to
the result in \cite[Theorem 1]{itw3} where the enrollment channel is
noiseless ($X=Y$) and the template rate can be arbitrarily large}.
Also, this characterization corresponds to the region given
by G\"unl\"u and Kramer \cite[Theorem 1]{onur} {with only one individual}.
It is easy to check this claim by just
setting $R_I = 0$.
Moreover, it is worthy mentioning that Kittichokechai and Caire \cite{kc} studied a similar model.
{They analyzed the model in which the enrollment channel is noise-free and the presence
of an adversary at the decoder is considered,
and characterized the capacity region by using two RVs as well}. In the case where there is no assumption of adversary,
it {can be confirmed} that the characterization in this paper {reduces to} their result \cite[Theorem 1]{kc} by similar arguments in the proof
of {(\ref{region333}) (Appendix A)}.

A numerical example of the rate region given by the right-hand side of
(\ref{theorem1}) where $R_S = 0$
is shown in Fig.\ \ref{tcp1}. This is a three-dimensional
figure of $R_I$ (z-axis) as a function of {$R_S$ (x-axis) and $R_J$ (y-axis),}
and the figure was plotted under the following settings.
We assume that alphabets $\mathcal{X}$, $\mathcal{Y}$, $\mathcal{Z}$, {and $\mathcal{U}$} are binary.
We fix source probability $P_X (0) = 0.5$ and
transition probability of channels $P_{Y|X}(0|0) = P_{Y|X}(1|1) = P_{Z|X}(0|0) = P_{Z|X}(1|1) = 0.9$.
The region below the curved surface of Fig.\ \ref{tcp1} is
the achievable rate region{, which is
a convex region, and it} stretches in the direction of blue arrow.
Fig.\ \ref{tcp2} shows a projection of the rate {region} onto the $R_JR_I$
plane and the colored area represents the achievable area of the rate pair {$(R_J,R_I)$}.
{Apparently, the template rate $R_J$ (storage space of the database) increases as the value of
the identification rate $R_I$ rises.}

\section{Proof of Theorem \ref{th1}} \label{sec4}
We take a standard information
theoretic approach; we divide the proof into the achievability (direct)
part {and} the converse part.

\subsection{Achievability {(Direct) Part}}

First, we fix $\delta > 0$ arbitrarily small, and a block length $n$.
We also fix test channels $P_{U|Y}$ and $P_{V|U}$.
We set\footnote{Due to the Markov chain $V-U-Z$, we have
$I(Z;U) - I(Z;V) = I(Z;UV) - I(Z;V) = I(Z;V) + I(Z;U|V) - I(Z;V)= I(Z;U|V)$.
In the proof, we use this fact without explanation.} $R_I = I(Z;V) - \delta$, $R_S = I(Z;U|V) -\delta$,
$R_J = I(Y;U) - I(Z;U) + I(Z;V) + 3\delta$, and $R_L = I(X;U) - I(Z;U) + I(Z;V) + 3\delta$.
We also set $M_I = 2^{nR_I}$,
{$M_S = 2^{nR_S}$}, and {$M_J = 2^{nR_J}$},
respectively.

\medskip
\noindent{\em Random Code Generation}:

Sequences $v^n_{m}$ are generated i.i.d.\ from $P_V$ for $m~{\in}~[1,N_V]$, where
$N_V = 2^{n\left(I(Y;V) + \delta\right)}$. For each $m$, sequences $u^n_{k|m}$ are generated from the memoryless
channel $P_{U^n|V^n=v_{m}^n}$ for $k~{\in}~[1,N_U]$, where $N_U = 2^{n\left(I(Y;U|V) + \delta\right)}$.
{Divide these sequences equally from the first index into
$N_B = {2^{n\left(I(Y;U|V) - I(Z;U|V) + 2\delta\right)}}$ bins. That is, the first bin contains $\{u^n_{1|m},\cdots,u^n_{M_S|m}\}$,
the second bin contains $\{u^n_{M_S+1|m},\cdots,u^n_{2M_S|m}\}$, and so on. Consequently, each bin contains
exactly $M_S$ codewords}. Bins are indexed by $b \in [1,N_B]$ and codewords inside a certain bin are indexed by
{$s \in \mathcal{S}$}. Without loss of generality, there exists a one-to-one mapping between $k$ and the pair $(b,s)$.

\medskip
\noindent{{\em Encoding (Enrollment)}}:

When encoder $f$ observes the bio-data sequence $y^n_i$,
{the} encoder {looks for $(m, k)$ such that $(y^n_i,v^n_{m},u^n_{k|m}) \in A_{\epsilon}^{(n)}(YVU)$}.
In case there are more than one such pairs, the encoder picks {one of them} uniformly at random.
Assume that the encoder found a corresponding pair {$(m,k) = (m(i), k(i))$}
satisfying the jointly typical condition above.
We set the template $j(i)=(m(i),b(i))$ and the secret data to be the corresponding codeword's index $s(i)$
{in bin} $b(i)$
\footnote{\scriptsize{{Since there is a one-to-one mapping between $k$ and $(b,s)$, we identify 
$k(i)$ with $(b(i),s(i))$.}}}.
$j(i)$ is stored at position $i$ in the database and
$s(i)$ is handed back to individual $i$.
If there do not exist such $m$ and $k$,
then we set $j(i) = {(1,1)}$ and $s(i) = 1$.

\medskip
\noindent{{\em Decoding (Identification)}}:

The decoder has access to {all records} in the database $\{(m(1), b(1)), \cdots (m(M_I), b(M_I))\}$.
When decoder $g$ sees $z^n$,
{the noisy version of identified individual sequence $x^n_w$},
it checks whether the codeword pair
$(v^n_{m(i)},u^n_{b(i),s|m(i)})$ is jointly typical with $z^n$ or not for all
$i \in \mathcal{I}$ {with} some
$s \in \mathcal{S}$, i.e.
$(z^n,v^n_{m(i)},u^n_{b(i),s|m(i)}) \in A_{\epsilon}^{(n)}(ZVU)$.
{If there exists a unique pair ${(i,s)}$ for which this condition holds,
then the decoder outputs
$(\widehat{w},\widehat{s(w)}) = {(i,s)}$
as the estimated index and secret data, respectively.
Otherwise, the decoder outputs {the index of the template $(1,1)$
as $\hat{w}$ and $\widehat{s(w)}=1$} if
{(i)} there does not exist such a pair $(i,s)$,
{(ii)} {such a pair $(i,s)$ exists} but there are some
$s' \neq {s}$ ($s' \in \mathcal{S}$) such that}
$(z^n,v^n_{{m(i)}},
u^n_{{b(i)},s'|{m(i)}}) \in A_{\epsilon}^{(n)}(ZVU)$ satisfies, or
{(iii)} {such a pair $(i,s)$ exists but} there are some $i' \neq {i}$ such that the pair
$(v^n_{m(i')},u^n_{b(i'),{s'}|m(i')})$
is jointly typical with $z^n$ for some {$s' \in \mathcal{S}$}.

\medskip
\noindent{{\em Analysis of {Error Probability}}}:

We evaluate the ensemble average of the error probability, 
where the average is taken over randomly chosen codebook $\mathcal{C}_n$,
which is defined as the set {$\{V_{m}^n, U_{k|m}^n:
m \in [1,N_V], k \in [1,N_U]\}$.}
Let the pair $(M(i),K(i)) = (M(i),B(i),S(i))$ denote the RVs corresponding to the index pair
$(m(i),k(i))=(m(i),b(i),s(i))$ of sequences $V_{m}^n$ and $U_{k|m}^n$
determined by the encoder for $Y^n_i$.
For individual {$W=w$}, possible event of errors occurs at the encoder is:

\begin{enumerate}[label={}]
	\item \hspace{-3mm}$\mathcal{E}_1$:~~\{$(Y^n_w,V^n_{m},U^n_{k|m}) \notin A_{\epsilon}^{(n)}(YVU)$ \\
        \hspace{6mm}  for all $m \in [1,N_V]$ and $k \in [1,N_U]$\},
        \medskip
        \\
\hspace{-8mm} and {those} at the decoder are:
	\item \hspace{-3mm}${\mathcal{E}_2}$:~~\{$(Z^n,V^n_{M(w)},U^n_{B(w),S(w)|M(w)}) \notin A_{\epsilon}^{(n)}(ZVU)$\},
        \item \hspace{-3mm}${\mathcal{E}_3}$:~~\{$\exists s' \neq S(w)$ s.\ t. \\
        \hspace{6mm}  $(Z^n,V^n_{M(w)},U^n_{B(w),s'|M(w)}) \in A_{\epsilon}^{(n)}(ZVU)$\},
        \item \hspace{-3mm}${\mathcal{E}_4}$:~~\{$\exists i'\neq w$ and $\exists s'$ s.\ t. \\
        \hspace{6mm}  $(Z^n,V^n_{M(i')},U^n_{B(i'),s'|M(i')}) \in A_{\epsilon}^{(n)}(ZVU)$\}.
\end{enumerate}
Then, the error probability for {$W=w$} can be bounded as
\begin{align}
&\max_{\substack{w \in \mathcal{I}}} \Pr\{(\widehat{W},\widehat{S(W)}) \neq (W,S(W))|W = w\} \nonumber \\
&~~~= \Pr\left\{\mathcal{E}_1\cup\mathcal{E}_2\cup\mathcal{E}_3\cup\mathcal{E}_4\right\} \nonumber \\
&~~~\overset{\mathrm{(a)}}\leq \Pr\left\{\mathcal{E}_1\right\} +
\Pr\left\{{\mathcal{E}_2|\mathcal{E}^c_1}\right\}+\Pr\left\{\mathcal{E}_3\right\} + \Pr\left\{\mathcal{E}_4\right\},
\end{align}
{where (a) follows because $\Pr\left\{\mathcal{E}_1,\mathcal{E}_2\right\} = \Pr\left\{\mathcal{E}_1\right\}
+\Pr\left\{\mathcal{E}_2\cap\mathcal{E}^c_1\right\} \le
\Pr\left\{\mathcal{E}_1\right\} + \Pr\left\{\mathcal{E}_2|\mathcal{E}^c_1\right\}$}.

$\Pr\{\mathcal{E}_1\}$ can be made smaller than $\delta$ for large enough $n$ by
{utilizing the covering lemma [12, Lemma 3.3]} because
{$\frac{1}{n}\log N_V = I(Y;V) + \delta~{>I(Y;V)}$} and {$\frac{1}{n}\log N_U = I(Y;U|V)+ \delta~{ >I(Y;U|V)}$}.
For $\Pr\left\{{\mathcal{E}_2|\mathcal{E}^c_1}\right\}$, 
it can also be made smaller than $\delta$ by the Markov lemma \cite[Lemma 15.8.1]{cover}.
{By applying the packing lemma {[12, Lemma 3.1]}, $\Pr\left\{{\mathcal{E}_3}\right\}$ and
$\Pr\left\{{\mathcal{E}_4}\right\}$ are arbitrarily
small for large enough $n$ since ${\frac{1}{n}\log M_S = I(Z;U|V)- \delta <I(Z;U|V)}$ and
${\frac{1}{n}\log M_I + \frac{1}{n}\log M_S = I(Z;U) -2\delta < I(Z;UV)}$, respectively.}

Therefore, {the ensemble average of} the error probability can be made that
\begin{align}
\max_{\substack{w \in \mathcal{I}}} \Pr\{(\widehat{W},\widehat{S(W)}) \neq (W,S(W))|W = {w}\}
\leq {4\delta} \label{error1}
\end{align}
for large enough $n$.

\medskip
\noindent{{\em Intermediate Steps}}:

{
We consider a {\em virtual} system, where a {\em partial} decoder {$g_i$} is employed,
for deriving the upper bound on the privacy-leakage rate. In this system, knowing index $i$ and seeing $Z^n_i$
({defined as the} output sequence of $X^n_i$ via $P_{Z|X}$), the partial decoder {$g_i$}
estimates only the secret data of individual $i$ as $\widehat{S(i)} = g_i(Z^n_i,J(i))$.
Note that this system is just for {analysis, and} the partial decoder is not actually used
during the decoding process.}

{For any given $i \in \mathcal{I}$}, {the partial decoder $g_i$ operates as follows: observing $z_i^n$ and the template
$j(i) = (m(i), b(i))$ in the database, it looks for $s \in \mathcal{S}$} such that
$(z^n_i,v^n_{m(i)},u^n_{b(i),s|m(i)}) \in A_{\epsilon}^{(n)}(ZVU)$. It sets $\widehat{s(i)} = s$ if there exists
{a unique} $s$. Otherwise, it outputs $\widehat{s(i)} = 1$.
The potential events of error probability for this case are ${\mathcal{E}_2}$ and ${\mathcal{E}_3}$.
{Letting} $P_e(i)$ be the error probability of {$g_i$, we readily see that}
\begin{align}
\hspace{-3mm} P_e(i) \le \Pr\{(\widehat{W},\widehat{S(W)})\neq (W,S(W))|W = i\}
\le {4\delta}, \label{error2}
\end{align}
where the middle term in (\ref{error2}) denotes
{the error probability of $g$ ({in the original BIS}) for individual $W = i$}.

The function of this partial decoder enables us to bound the following conditional entropy
\begin{align}
H(S(i)|Z^n_i,{J(i)},\mathcal{C}_n) \overset{\mathrm{(b)}}\le H(S(i)|\widehat{S(i)})
\overset{\mathrm{(c)}}\le n\delta_n, \label{fano1}
\end{align}
where
\begin{enumerate}[label=(\alph*)]
	\setcounter{enumi}{1}
	\item follows because conditioning reduces entropy,
        \item {follows because {Fano's inequality and (\ref{error2}) are} applied, and
        ${\delta_n}=\frac{1}{n}\left(1 + 4\delta\log M_S\right)$}.
\end{enumerate}

\begin{lemma} (Kittichokechai et al.\ \cite{kitti}) \label{keylem}

Assume that $(X^n,Y^n,U^n)$ are {$\epsilon$-strongly typical} with high
probability\footnote{\scriptsize{It means that $\Pr\{(X^n,Y^n,U^n)\in A^{(n)}_{\epsilon}(XYU)\} \rightarrow 1$ as $n\rightarrow \infty$,
where $A^{(n)}_{\epsilon}(XYU)$ denotes the set of $\epsilon$-strongly typical sequences.}}. Then, it holds that
\begin{align}
\frac{1}{n}H(Y^n|U^n,\mathcal{C}_n) \leq H(Y|U) + \delta'_n, \label{ynun} \\
\frac{1}{n}H(Y^n|X^n,U^n,\mathcal{C}_n) \leq H(Y|X,U) + {\delta'_n}, \label{ynxnun}
\end{align}
where $\delta'_n $ {is a positive value {satisfying}
${\delta'_n} \downarrow 0$}.
\end{lemma}
\noindent{{(Proof)~~~~The {proofs} can be found in
\cite[Appendix C]{kitti}}}.
\qed

\begin{lemma} \label{keylem111}

{For any $i \in \mathcal{I}$}, it holds that
\begin{align}
\frac{1}{n}H(Y^n_i|J(i),S(i),\mathcal{C}_n)\leq H(Y|U) + {\delta'_n}, \label{lemma2222}
\end{align}
where $\delta'_n > 0$ {and ${\delta'_n} \downarrow 0$}.
\end{lemma}
\noindent{{(Proof)}}~~~~The proof is provided in Appendix B.
\qed

{Due to the fact that we set $M_S = 2^{nR_S}$ and $M_J = 2^{nR_J}$,
the following inequalities hold
\begin{align}
    \frac{1}{n}H(S(i)|\mathcal{C}_n) &\le R_S = I(Z;U|V)-\delta,  \label{sicn} \\
    \frac{1}{n}H(J(i)|\mathcal{C}_n) &\le R_J = I(Y;U)-I(Z;U|V)+3\delta \label{jicn}
\end{align}
with equality when $S(i)$ and $J(i)$ are uniformly distributed on $\mathcal{S}$ and $\mathcal{J}$, respectively,
for {any} codebook $\mathcal{C}_n$.}

{Hereafter, we shall check the bounds of {identification}, secrecy, secrecy-leakage, template, and
privacy-leakage rates averaged over randomly chosen codebook $\mathcal{C}_n$.
In the following analyses, the index $i$ is arbitrarily fixed on $\mathcal{I}$
since we need to show that all conditions in Definition 1 are satisfied.}

\medskip
\noindent{{\em {Analyses} of {{Identification and} Template Rates}}}:

From the parameter settings of achievability scheme,
it is straight-forward that the {conditions (\ref{b}) and (\ref{d}) hold}.

\medskip
\noindent{{\em Analysis of Secrecy Rate}}:

The secrecy rate can be evaluated as follows:
\begin{align}
\hspace{-6mm}\frac{1}{n}H(S(i)|{\mathcal{C}_n}) &= \frac{1}{n}\Big\{H(Y^n_i,J(i),S(i)|{\mathcal{C}_n})
- H(J(i)|S(i),{\mathcal{C}_n}) \nonumber  \\
&~~~~~~~~~~~- H(Y^n_i|J(i),S(i),{\mathcal{C}_n})\Big\} \nonumber  \\
&\overset{\mathrm{(d)}}\geq \frac{1}{n}\Big\{H(Y^n_i) - H(J(i)|\mathcal{C}_n)\nonumber  \\
&~~~~~~~~~~~-H(Y^n_i|J(i),S(i),{\mathcal{C}_n})\Big\} \nonumber  \\
&\overset{\mathrm{(e)}}\geq H(Y) - {(I(Y;U) - I(Z;U) + I(Z;V)+3\delta)} \nonumber  \\
&~~~~~~~~~~~- (H(Y|U) + {\delta'_n}) \nonumber  \\
&= I(Z;U) - I(Z;V) - 3\delta - {\delta'_n} \nonumber  \\
&\overset{\mathrm{(f)}}= R_S -2\delta - {\delta'_n}, \label{RHV}
\end{align}
where
\begin{enumerate}[label=(\alph*)]
	\setcounter{enumi}{3}
        \item holds because $(J(i),S(i))$ is a function of $Y^n_i$,
        \item {follows because (\ref{jicn}) and Lemma 2 are applied},
        \item holds because we set $R_S = I(Z;U) - I(Z;V)-\delta$.
\end{enumerate}

\medskip
\noindent{{\em Analysis of Secrecy-Leakage}}:

The amount of leaked information about $S(i)$ from $J(i)$
can be expanded as
\begin{align}
\hspace{-5mm}\frac{1}{n}I(&J(i);S(i)|{\mathcal{C}_n}) \nonumber  \\
=&\frac{1}{n}\{H(S(i)|{\mathcal{C}_n}) + H(J(i)|{\mathcal{C}_n}) - H(Y^n_i,J(i),S(i)|{\mathcal{C}_n}) \nonumber  \\
&~~~~~+H(Y^n_i|J(i),S(i),{\mathcal{C}_n})\} \nonumber  \\
=&\frac{1}{n}H(S(i)|{\mathcal{C}_n}) + \frac{1}{n}H(J(i)|{\mathcal{C}_n}) - \frac{1}{n}H(Y^n_i) \nonumber  \\
&~~~~~+\frac{1}{n}H(Y^n_i|J(i),S(i),{\mathcal{C}_n}) \nonumber  \\
\overset{\mathrm{(g)}}\leq& I(Z;U|V) - \delta +  I(Y;U) - I(Z;U|V) + 3\delta - H(Y)  \nonumber  \\
&~~~~~+ H(Y|U) +{\delta'_n} \nonumber  \\
=&2\delta + {\delta'_n}, \label{IJS}
\end{align}
where {(g)} follows because {(\ref{sicn}), (\ref{jicn}), and Lemma 2 are applied}.

\medskip
\noindent{{\em Analysis of Privacy-Leakage {Rate}}}:

{In view of (\ref{dd})}, we start by expanding the
{privacy-leakage rate $\frac{1}{n}I(X^n_i;J(i)|\mathcal{C}_n)$ as}
\begin{align}
\hspace{-5mm}\frac{1}{n}I(X^n_i;J(i)|\mathcal{C}_n)
&= \frac{1}{n}H(J(i)|\mathcal{C}_n) - \frac{1}{n}H(J(i)|X^n_i,\mathcal{C}_n) \nonumber \\
&\leq I(Y;U) - I(Z;U) + I(Z;V) + 3\delta \nonumber  \\
&~~~~~ - \frac{1}{n}H(J(i)|X^n_i,\mathcal{C}_n). \label{21244}
\end{align}
{where the inequality in (\ref{21244}) holds due to (\ref{jicn})}. Next, {let us} focus solely on the {conditional entropy} in (\ref{21244}).
It can {be evaluated} as
\begin{align}
\hspace{-5mm}\frac{1}{n}H(&J(i)|X^n_i,\mathcal{C}_n) \nonumber \\
&= \frac{1}{n}H(Y^n_i,J(i)|X^n_i,\mathcal{C}_n) - \frac{1}{n}H(Y^n_i|J(i),X^n_i,\mathcal{C}_n) \nonumber \\
&\overset{\mathrm{(h)}}= \frac{1}{n}H(Y^n_i|X^n_i,\mathcal{C}_n) - \frac{1}{n}H(Y^n_i|M(i),B(i),X^n_i,\mathcal{C}_n) \nonumber \\
&\overset{\mathrm{(i)}}= H(Y|X) - \frac{1}{n}H(Y^n_i|M(i),B(i),S(i),X^n_i,\mathcal{C}_n) \nonumber \\
&~~~~~-\frac{1}{n}I(S(i);Y^n_i|M(i),B(i),X^n_i,\mathcal{C}_n) \nonumber \\
&\ge H(Y|X) - \frac{1}{n}H(Y^n_i|M(i),B(i),S(i),X^n_i,\mathcal{C}_n) \nonumber \\
&~~~~~-\frac{1}{n}H(S(i)|M(i),B(i),X^n_i,\mathcal{C}_n) \nonumber \\
&\overset{\mathrm{(j)}}= H(Y|X) - \frac{1}{n}H(Y^n_i|M(i),B(i),S(i),U^n_i,X^n_i,\mathcal{C}_n) \nonumber \\
&~~~~~-\frac{1}{n}H(S(i)|M(i),B(i),X^n_i,{Z^n_i},\mathcal{C}_n) \nonumber \\
&\overset{\mathrm{(k)}}\ge H(Y|X) - \frac{1}{n}H(Y^n_i|U^n_i,X^n_i,\mathcal{C}_n) \nonumber \\
&~~~~~-\frac{1}{n}H(S(i)|M(i),B(i),{Z^n_i},\mathcal{C}_n) \nonumber \\
&\overset{\mathrm{(l)}}\ge H(Y|X) - H(Y|X,U) - (\delta_n + {\delta'_n}) \nonumber \\
&{= I(Y;U|X) - (\delta_n + {\delta'_n})} \nonumber \\
&\overset{\mathrm{(m)}}= H(U|X) - H(U|Y) - (\delta_n + {\delta'_n}), \label{1234567899}
\end{align}
where
\begin{enumerate}[label=(\alph*)]
	\setcounter{enumi}{7}
        \vspace{-2mm}
\item follows since $J(i)$ is a function of $Y^n_i$ and we have $J(i) = (M(i),B(i))$,
\item follows because {$Y^n_i$ and $X^n_i$ are independent of $\mathcal{C}_n$},
\item follows because {$U^n_{B(i),S(i)|M(i)}$} is denoted by $U^n_i$ and
it is a function of the tuple $(M(i),B(i),S(i))$ for the second term, and the Markov chain
$S(i)-(M(i),B(i),X^n_i)-{Z^n_i}$ holds for a given codebook in the last term,
\item follows because conditioning reduces entropy,
\item {follows as (\ref{ynxnun}) in Lemma 1 and Fano's inequality in (\ref{fano1}) are applied},
\item {holds since we have $H(U|Y,X)=H(U|Y)$ by the Markov chain $U-Y-X$}.
\end{enumerate}

From (\ref{21244}) and (\ref{1234567899}), we obtain
\begin{align}
\hspace{-5mm}\frac{1}{n}I(X^n_i;J(i)|\mathcal{C}_n) &\leq H(U) - H(U|Y) - I(Z;U) + I(Z;V) \nonumber \\
&~~~+ H(U|Y) - H(U|X) + 3\delta + {\delta_n+ {\delta'_n}} \nonumber \\
&\leq I(X;U) - I(Z;U) + I(Z;V) \nonumber \\
&~~~+ 3\delta + {\delta_n+{\delta'_n}} \nonumber \\
&\leq {R_L + \delta}
\label{IJX}
\end{align}
for all sufficiently large $n$.

Finally, {with a sufficiently small $\delta$ and by} applying the selection lemma
\cite[Lemma 2.2]{BB} to all results shown above (i.e., Eqs.\ (17), (25), (26), and (29)), there exists
a codebook satisfying all the conditions in Definition 1 for all large enough $n$.
\qed

\subsection{Converse Part}

{For the converse proof, we consider a more relaxed case where} identified individual index $W$
is {\em uniformly} distributed over $\mathcal{I}$ and (\ref{a}) is replaced with
the average {error criterion}
\begin{align}
\Pr\{({\widehat{W}},\widehat{S(W)})\neq (W,S(W))\} \leq \delta. \label{average}
\end{align}
{We shall} show that the capacity region, which is not smaller than the original one {$\mathcal{R}$},
is contained in the right-hand side of
(\ref{theorem2}).

We assume that a rate tuple $(R_I,R_S,R_J,R_L)$ is achievable so that
there exists a pair of encoder and decoder $(f,g)$ such that all conditions in
Definition \ref{def1} with replacing (\ref{a}) by (\ref{average}) are satisfied for any
$\delta > 0$ and large enough $n$. 

Here, we provide {other key} lemmas used in this part.
For $t \in [1,n]$, we define auxiliary RVs $U_t$ and $V_t$ as
$
U_t = (Z^{t-1},J(W),S(W))
$
and
$
V_t = (Z^{t-1},J(W))
$, respectively. We denote a sequence of RVs
{$X^n_W  = (X_1(W) ,\cdots,X_n(W))$ and}
$Y^n_W  = (Y_1(W) ,\cdots,Y_n(W))$.

\begin{lemma} \label{markov122}

{The following Markov chains hold}
\begin{align}
Z^{t-1}-(Y^{t-1}(W),J(W),S(W))-Y_t(W), \label{zyyt} \\
Z^{t-1}-(X^{t-1}(W),J(W),S(W))-X_t(W). \label{zxxt}
\end{align}
\end{lemma}
\noindent{{(Proof)}}~~~~The {proofs are} given in Appendix C.
\qed

\begin{lemma} \label{leema}
There {exist} some RVs $U$ and $V$ which satisfy $Z-X-Y-U-V$ and
\begin{align}
\sum_{\substack{ t=1 }}^nI(Z_{t};V_t) &= nI(Z;V), \label{ztvt} \\
\sum_{\substack{ t=1 }}^nI(Z_{t};U_t) &= nI(Z;U), \label{ztut} \\
{\sum_{\substack{ t=1 }}^nI(Y_{t}(W);U_t)} &= nI(Y;U), \label{ytut} \\
\sum_{\substack{ t=1 }}^nI(X_{t}(W);U_t) &= nI(X;U). \label{xtut}
\end{align}
\end{lemma}

\noindent{(Proof)}~~~~The {proofs are} provided in Appendix D.
\qed

In the {subsequent} analyses, we fix auxiliary RVs $U$ and $V$ specified in Lemma \ref{leema}.

\medskip
\noindent{{\em Analysis of Identification Rate:}}

Again note that we are considering the case where $W$ is
uniformly distributed in the converse part, {and we have}
\begin{align}
\log M_I &= H(W) \nonumber \\
&= H(W|\mathcal{J}_{M_I},Z^n) + I(W;\mathcal{J}_{M_I},Z^n) \nonumber \\
&\overset{\mathrm{(a)}}= H(W|\mathcal{J}_{M_I},Z^n,\widehat{W},\widehat{S(W)}) + I(W;\mathcal{J}_{M_I},Z^n) \nonumber \\
&\overset{\mathrm{(b)}}\leq H(W|\widehat{W},\widehat{S(W)}) + I(W;\mathcal{J}_{M_I},Z^n) \nonumber \\
&\leq H(W,S(W)|\widehat{W},\widehat{S(W)}) + I(W;\mathcal{J}_{M_I},Z^n), \label{123456}
\end{align}
where
\\
(a) holds because $(\widehat{W},\widehat{S(W)})$ is function of
$\mathcal{J}_{M_I}$ and $Z^n$, \\
(b) follow{s} because conditioning reduces entropy.
\\
Continue bounding the second term in (\ref{123456}),
\begin{align}
I(W;\mathcal{J}_{M_I},Z^n) &= I(W;\mathcal{J}_{M_I}) + I(W;Z^n|\mathcal{J}_{M_I}) \nonumber \\
&\overset{\mathrm{(c)}}= I(W;Z^n|\mathcal{J}_{M_I}) \nonumber \\
&= H(Z^n|\mathcal{J}_{M_I}) - H(Z^n|\mathcal{J}_{M_I},W)\nonumber \\
&\overset{\mathrm{(d)}}= H(Z^n|J(W)) - H(Z^n|J(W),W)\nonumber \\
&\overset{\mathrm{(e)}}\leq H(Z^n) - H(Z^n|J(W),W)\nonumber \\
&= H(Z^n) - H(Z^n|J(W)) \nonumber \\
&= \sum_{\substack{ t=1 }}^n \Big\{H(Z_t) - H(Z_t|Z^{t-1},J(W)) \Big\} \nonumber \\
&= \sum_{\substack{ t=1 }}^n I(Z_t;V_t) \overset{\mathrm{(f)}}= nI(Z;V), \label{last_step}
\end{align}
where
\begin{enumerate}[label=(\alph*)]
\setcounter{enumi}{2}
\item follow{s} because $W$ is independent of other RVs,
\item follows because only $J(W)$ is possibly dependent on $Z^n$,
\item follows because conditioning reduces entropy,
\item follows because of (\ref{ztvt}) in Lemma \ref{leema}.
\end{enumerate}
Thus, from (\ref{b}), (\ref{123456}), (\ref{last_step}), {and Fano's inequality as in (\ref{fano1})}, we obtain
\begin{align}
R_I \le I(Z;V) + {\delta} + \delta_n, \label{rilast}
\end{align}
where $\delta_n = \frac{1}{n}(1+\delta\log M_IM_S)$
and{\footnote{{Willems et al.\ [2] characterized the identification capacity of
the system, where the decoder estimates only the user index, and showed that
$\frac{1}{n}\log M_I \le I(Y;Z) + \delta$ for all sufficiently large $n$. Since the constraints imposed on the
system addressed in this paper are more rigorous than the ones in [2], it is trivial that $\frac{1}{n}\log M_I$
for this system cannot be larger than $I(Y;Z) + \delta$. Moreover, it holds that
$\frac{1}{n}\log M_S \le \log|\mathcal{Y}|$ because $S(i)$ is a function of $Y^n_i$. Therefore, for large enough
$n$, we have that  $\delta_n = \frac{1}{n} + \frac{\delta }{n}\log M_IM_S \le \frac{1}{n} + \delta(\log|\mathcal{Y}||\mathcal{Z}|+\delta)$,
and it converges to zero when $n \rightarrow \infty$ and $\delta \downarrow 0$.}}}
{$\delta_n \downarrow 0$} as $n \rightarrow \infty$
{and $\delta \downarrow 0$}.

\medskip
\noindent{{\em Analysis of Secrecy Rate:}}

This analysis is similar to {the analysis of identification rate},
which we have already seen above. We begin by considering the entropy of
secret data as follows:
\begin{align}
H(S(W))&= H(S(W)|\mathcal{J}_{M_I},Z^n) + I(S(W);\mathcal{J}_{M_I},Z^n) \nonumber \\
&= H(S(W)|\mathcal{J}_{M_I}Z^n,\widehat{W},\widehat{S(W)}) \nonumber \\
&~~~~~+ I(S(W);\mathcal{J}_{M_I},Z^n) \nonumber \\
&\leq H(S(W)|\widehat{W},\widehat{S(W)}) + I(S(W);\mathcal{J}_{M_I},Z^n) \nonumber \\
&\leq H(W,S(W)|\widehat{W},\widehat{S(W)}) + I(S(W);\mathcal{J}_{M_I},Z^n) \nonumber \\
&= H(W,S(W)|\widehat{W},\widehat{S(W)}) + I(S(W);\mathcal{J}_{M_I})\nonumber \\
&~~~~~+ I(S(W);Z^n|\mathcal{J}_{M_I}) \nonumber \\
&\overset{\mathrm{(g)}}= H(W,S(W)|\widehat{W},\widehat{S(W)}) + I(S(W);J(W))\nonumber \\
&~~~~~+ I(S(W);Z^n|J(W)), \label{123456sw}
\end{align}
where (g) follows because bio-data sequence of each individual is generated
independently so
only $J(W),S(W)$, and $Z^n$ are possibly dependent on each other.
\\
For the third term in (\ref{123456sw}),
\begin{align}
I(S(W)&;Z^n|J(W)) \nonumber \\
&= H(Z^n|J(W)) - H(Z^n|J(W),S(W)) \nonumber \\
&= H(Z^n) - H(Z^n|J(W),S(W)) \nonumber \\
&~~~~~- (H(Z^n) - H(Z^n|J(W))) \nonumber \\
&\overset{\mathrm{(h)}}= \sum_{\substack{ t=1 }}^n \Big\{H(Z_t) - H(Z_t|Z^{t-1},J(W),S(W))\Big\} \nonumber \\
&~~~~~-\sum_{\substack{ t=1 }}^n \Big\{H(Z_t) - H(Z_t|Z^{t-1},J(W))\Big\} \nonumber \\
&= \sum_{\substack{ t=1 }}^n \Big\{I(Z_t;U_t) - I(Z_t;V_t)\Big\} \nonumber \\
&\overset{\mathrm{(i)}}= n(I(Z;U)-I(Z;V)), \label{last_stepsw}
\end{align}
where
\begin{enumerate}[label=(\alph*)]
\setcounter{enumi}{7}
\item holds because each symbol of $Z^n$ is i.i.d,
\item holds due to (\ref{ztvt}) and (\ref{ztut}) in Lemma \ref{leema}.
\end{enumerate}
Therefore, from (\ref{c}), (\ref{shiki4}), (\ref{123456sw}), (\ref{last_stepsw}),
{and Fano's inequality}, we have
\begin{align}
R_S \le I(Z;U)-I(Z;V) + {2\delta} + {\delta_n}. \label{rslast}
\end{align}

\newpage
\noindent{{\em Analysis {of} Template Rate:}}

It follows from (\ref{d}) that 
\begin{align}
\hspace{-5mm}n(&R_J + \delta) \nonumber \\
&\geq \log M_J \geq H(J(W)) \nonumber \\
& = I(Y^n_W;J(W)) \nonumber \\
& = I(Y^n_W;J(W),S(W),Z^n) - I(Y^n_W;Z^n|J(W)) \nonumber \\
&~~~~~- I(Y^n_W;S(W)|J(W),Z^n). \label{rjaaa}
\end{align}
Now {let us} focus on each term in (\ref{rjaaa}) separately. For the first term,
\begin{align}
\hspace{-8mm}I&(Y^n_W;J(W),S(W),Z^n) \nonumber \\
& = I(Y^n_W;J(W),S(W)) + I(Y^n_W;Z^n|J(W),S(W)) \nonumber \\
& = \sum_{\substack{ t = 1 }}^n\Big\{H(Y_t(W)) - H(Y_t(W)|Y^{t-1}(W),J(W),S(W))\Big\} \nonumber \\
&~~~~~ + H(Z^n|J(W),S(W)) - H(Z^n|J(W),S(W),Y^n_W)\nonumber \\
& \overset{\mathrm{(j)}}= \sum_{\substack{ t = 1 }}^n \Big\{H(Y_t(W))\nonumber \\
&~~~~~ - H(Y_t(W)|Z^{t-1},Y^{t-1}(W),J(W),S(W))\Big\} \nonumber \\
&~~~~~ + \sum_{\substack{ t = 1 }}^n H(Z_t|Z^{t-1},J(W),S(W)) - H(Z^n|Y^n_W)\nonumber \\
& \overset{\mathrm{(k)}}\ge \sum_{\substack{ t = 1 }}^n \Big\{H(Y_t(W)) - H(Y_t(W)|Z^{t-1},J(W),S(W))\Big\} \nonumber \\
&~~~~~ + \sum_{\substack{ t = 1 }}^n H(Z_t|U_t) - nH(Z|Y)\nonumber \\
& = \sum_{\substack{ t = 1 }}^n \Big\{I(Y_t(W);U_t)+ H(Z_t|U_t)\Big\} - nH(Z|Y), \label{rjbb}
\end{align}
where
\begin{enumerate}[label=(\alph*)]
\setcounter{enumi}{9}
\item holds from (\ref{zyyt}) in Lemma \ref{markov122} and $(S(W),J(W))$ is a function of $Y^n_W$,
\item follows because conditioning reduces entropy.
\end{enumerate}
For the second term,
\begin{align}
\hspace{-5mm}I(Y^n_W;Z^n|J(W)) &= H(Z^n|J(W)) - H(Z^n|J(W),Y^n_W) \nonumber \\
&=\sum_{\substack{ t = 1 }}^n H(Z_t|Z^{t-1},J(W)) - H(Z^n|Y^n_W) \nonumber \\
&=\sum_{\substack{ t = 1 }}^n H(Z_t|V_t) - nH(Z|Y). \label{rjbb1}
\end{align}
For the last one,
\begin{align}
\hspace{-5mm} I(Y^n_W;S(W)|J(W),Z^n) &\le H(S(W)|J(W),Z^n) \nonumber \\
&= H(S(W)|\mathcal{J}_{M_I},Z^n) \nonumber \\
&= H(S(W)|\mathcal{J}_{M_I},Z^n,\widehat{W},\widehat{S(W)}) \nonumber \\
& \overset{\mathrm{(l)}}\le H(S(W)|\widehat{W},\widehat{S(W)}) \nonumber \\
& \overset{\mathrm{(m)}}\le n\delta_n, \label{rjbb2}
\end{align}
where
\begin{enumerate}[label=(\alph*)]
\setcounter{enumi}{11}
\item follows because conditioning reduces entropy,
\item follows due to {Fano's inequality}.
\end{enumerate}
Finally, substituting
(\ref{rjbb})--(\ref{rjbb2}) into (\ref{rjaaa}), the last {terms} in (\ref{rjbb}) and (\ref{rjbb1}) cancel out
each other, and we obtain
\begin{align}
R_J &+ \delta \nonumber \\
&\ge \frac{1}{n}\sum_{\substack{ t = 1 }}^n \left\{I(Y_t(W);U_t) + H(Z_t|U_t) - H(Z_t|V_t)\right\}
-{\delta_n} \nonumber \\
&= \frac{1}{n}\sum_{\substack{ t = 1 }}^n \left\{I(Y_t(W);U_t) - I(Z_t;U_t) + I(Z_t;V_t)\right\}
-{\delta_n} \nonumber \\
&=I(Y;U) - I(Z;U) + I(Z;V) - {\delta_n}, \label{rjlast}
\end{align}
where (\ref{rjlast}) follows due to (\ref{ztvt})--(\ref{ytut}) in Lemma \ref{leema}.

\medskip
\noindent{{\em Analysis of Privacy-Leakage Rate:}}

From (\ref{dd}), it follows that
\begin{align}
n(R_L + \delta) &\geq \max_{\substack{w \in \mathcal{I}}}{I(X^n_w;J(w))} \nonumber \\
& \ge I(X^n_W;J(W)|W) = I(X^n_W;J(W)) \nonumber \\
& = I(X^n_W;J(W),S(W),Z^n) - I(X^n_W;Z^n|J(W)) \nonumber \\
&~~~~~- I(X^n_W;S(W)|J(W),Z^n). \label{rlaaa}
\end{align}
Likewise in the analysis of template rate, {let us} focus on each term in (\ref{rlaaa}) separately. For the first term,
\begin{align}
&I(X^n_W;J(W),S(W),Z^n) \nonumber \\
& = I(X^n_W;J(W),S(W)) + I(X^n_W;Z^n|J(W),S(W)) \nonumber \\
& \overset{\mathrm{(n)}}\ge I(X^n_W;J(W),S(W)) + H(Z^n|J(W),S(W)) \nonumber \\
&~~~~~ - H(Z^n|J(W),X^n_W)\nonumber \\
& \overset{\mathrm{(o)}}\ge \sum_{\substack{ t = 1 }}^n \Big\{H(X_t(W))\nonumber \\
&~~~~~- H(X_t(W)|Z^{t-1},X^{t-1}(W),J(W),S(W))\Big\} \nonumber \\
&~~~~~ + \sum_{\substack{ t = 1 }}^n H(Z_t|Z^{t-1},J(W),S(W)) - H(Z^n|J(W),X^n_W)\nonumber \\
& \overset{\mathrm{(p)}}\ge \sum_{\substack{ t = 1 }}^n \Big\{H({X_t(W)}) - H({X_t(W)}|Z^{t-1},J(W),S(W))\Big\} \nonumber \\
&~~~~~ + \sum_{\substack{ t = 1 }}^n H(Z_t|U_t) - H(Z^n|J(W),X^n_W)\nonumber \\
& = \sum_{\substack{ t = 1 }}^n \Big\{I({X_t(W)};U_t)+ H(Z_t|U_t)\Big\} - H(Z^n|J(W),X^n_W), \label{rlbb}
\end{align}
where
\begin{enumerate}[label=(\alph*)]
\setcounter{enumi}{13}
\item follows because conditioning reduces entropy,
\item holds from (\ref{zxxt}) in Lemma \ref{markov122},
\item follows because conditioning reduces entropy.
\end{enumerate}
For the second term,
\begin{align}
I&(X^n_W;Z^n|J(W)) \nonumber \\
&= H(Z^n|J(W)) - H(Z^n|J(W),X^n_W) \nonumber \\
&=\sum_{\substack{ t = 1 }}^n H(Z_t|Z^{t-1},J(W)) - H(Z^n|J(W),X^n_W) \nonumber \\
&=\sum_{\substack{ t = 1 }}^n H(Z_t|V_t) - H(Z^n|J(W),X^n_W), \label{rlbb1}
\end{align}
and the last term can be bounded by the same quantity as seen in (\ref{rjbb2}):
\begin{align}
I(X^n_W;S(W)|J(W),Z^n) &\le n{\delta_n}. \label{rlbb2}
\end{align}
Finally, substituting
(\ref{rlbb})--(\ref{rlbb2}) into (\ref{rlaaa}) and {taking} similar steps as in (\ref{rjlast}), we obtain
\begin{align}
R_L &+ \delta \nonumber \\
& \ge \frac{1}{n}\sum_{\substack{ t = 1 }}^n \left\{I(X_t(W);U_t) - I(Z_t;U_t) + I(Z_t;V_t)\right\} - {\delta_n} \nonumber \\
&= I(X;U) - I(Z;U) + I(Z;V) - {\delta_n}, \label{rllast}
\end{align}
where (\ref{rllast}) follows due to (\ref{ztvt}), (\ref{ztut}), and (\ref{xtut}) in Lemma \ref{leema}.

Eventually, letting $n \rightarrow \infty$ and $\delta \downarrow 0$ in
(\ref{rilast}), (\ref{rslast}), (\ref{rjlast}), and (\ref{rllast}),
we can see that the capacity region is contained in the right-hand side of (\ref{theorem2}).

{To complete} the proof of Theorem \ref{th1}, {we discuss} the bounds on
the cardinalities of auxiliary RVs.
For proving the bound on the cardinality of alphabet $\mathcal{U}$ in the region $\mathcal{A}_1$
{(cf.\ (\ref{theorem1}))},
we use the support lemma in \cite[Appendix C]{GK} to show that RV $U$ should have $|\mathcal{Y}|-1$
elements to preserve $P_{Y}$ and add three more elements to preserve $H(Z|U)$, $H(Y|U)$, and $H(X|U)$.
This implies that it suffices to take $|\mathcal{U}| \le |\mathcal{Y}|+2$
for preserving $\mathcal{A}_1$.
Similarly, to bound the cardinalities of alphabets $\mathcal{U}$ and $\mathcal{V}$
in the region $\mathcal{A}_2$ {(cf.\ (\ref{theorem2}))}, we also utilize the same lemma to show that
$|\mathcal{V}| \le |\mathcal{Y}|+3$ and $|\mathcal{U}| \le (|\mathcal{Y}| + 2)(|\mathcal{Y}| + 3)$
suffice to preserve $P_{Y}$, $H(Z|V)$, $H(Z|U)~(= H(Z|U,V))$, $H(Y|U)$, and $H(X|U)$.
\qed

\section{Conclusions and Future Works} \label{sec5}
In this paper, we deployed a method using two auxiliary RVs to characterize
the capacity region of identification, secrecy, template, {and} privacy-leakage
rates in the BIS.
{We demonstrated that the characterization using two auxiliary RVs reduce to
the one using only an auxiliary RV.}
Compared to the model proposed in \cite{tuncel} and \cite{itw3}, what we newly imposed on our model are:
\begin{itemize}
	\item treating a noisy channel in the enrollment phase,
	\item considering a scheme of both compressing template (as in \cite{tuncel} and \cite{vy})
        and protecting privacy (as in \cite{itw3}),
	\item analyzing the capacity region provided that the prior distribution of an identified individual is unknown.
\end{itemize}

As special cases, it can be checked that our characterization reduces to the one in \cite[Theorem 1]{itw3}
where the enrollment channel is noiseless and there is no constraint on the template rate, and
it also coincides with the {one} derived by G\"unl\"u and Kramer \cite[Theorem 1]{onur} {where there is
only one individual}.
{In a slightly different model in which the secret key is chosen independently of the bio-data sequences,
known as chosen-secret BIS model \cite{itw3},\cite{onur}, the capacity region has not been discussed in this paper.
However, it can be characterized via similar arguments for proving Theorem 1 by just adding a one-time pad operation}.
For the future works, as we have seen in Remark \ref{remark2222} about the relation between $\mathcal{A}_1$
and $\mathcal{A}_2$, this is a positive hint that Theorem \ref{th1} can be reproved by a scheme {using}
only one auxiliary RV and now {this task is under way}. {We also} plan
to analyze the capacity regions of the BIS under strong secrecy criterion regarding secrecy-leakage.

\appendices
\section{\em Proof of Equation (\ref{region333})}

In the proof, we show the equivalence of $\mathcal{A}_1$ and $\mathcal{A}_2$
by removing the cardinality bounds of auxiliary RVs $U$ and $V$ from
the two regions. Once the equivalence without the cardinality bounds is established,
the cardinality bounds follow from the standard arguments (cf.\ \cite[Appendix C]{GK}).

It is obvious that $\mathcal{A}_2 \subseteq \mathcal{A}_1$,
so we shall show that {$\mathcal{A}_2 \supseteq \mathcal{A}_1$}. We assume that
$(R_I,R_S,R_J,R_L) \in \mathcal{A}_1$, {meaning that $(R_I,R_S,R_J,R_L)$ satisfies all conditions
in (\ref{theorem1})} for some $P_{U|Y}$.
{Especially}, we have $R_I+R_S \le I(Z;U)$.
We choose the test channel $P_{V|U}$ satisfying
that
\begin{align}
R_I = I(Z;V).
\end{align}
Such $P_{V|U}$ always exists since $I(Z;U) \ge I(Z;V) \ge 0$
{and $I(Z;V)$ is a continuous function of $P_{V|U}$}.
Under that condition, it is easy to check that {$(R_I,R_S,R_J,R_L)$}
is also an element lying in the region $\mathcal{A}_2$.
\qed

\section{\em Proof of Lemma \ref{keylem111}}
In \cite{itw3}, a similar result of this lemma is used without the proof. Here, we will provide
{a} proof for readers' sake.

Note that $J(i) = (M(i),B(i))$. We start by considering the conditional entropy
in the {left}-hand side of (\ref{lemma2222}) {as}
\begin{align}
\textstyle \frac{1}{n}H(Y^n_i|J(i),S(i),\mathcal{C}_n) &= \textstyle \frac{1}{n}H(Y^n_i|M(i),B(i),S(i),\mathcal{C}_n) \nonumber \\
&\overset{\mathrm{(a)}}= \textstyle \frac{1}{n}H(Y^n_i|M(i),B(i),S(i),U^n_i,\mathcal{C}_n) \nonumber \\
&\overset{\mathrm{(b)}}\le \textstyle \frac{1}{n}H(Y^n_i|U^n_i,\mathcal{C}_n) \nonumber \\
&\overset{\mathrm{(c)}}\le H(Y|U) + \delta'_n
\end{align}
where
\begin{enumerate}[label=(\alph*)]
\setcounter{enumi}{0}
\item holds because we denote ${U^n_{B(i),S(i)|M(i)}}$ as $U^n_i$ for simplicity and
the tuple $(M(i),B(i),S(i))$ determines $U^n_i$ for a given codebook,
\item follows because conditioning reduces entropy,
\item follows because $Y^n_i$ and $U^n_i$ are jointly typical with high probability and (\ref{ynun}) in Lemma \ref{keylem}
is applied.
\end{enumerate}
\qed

\section{\em Proof of Lemma \ref{markov122}}
First, we prove that (\ref{zyyt}) holds. The joint distribution among
$Z^{t-1},Y^{t}(W),J(W)$, and $S(W)$ can be developed as
\begin{align*}
&P_{Z^{t-1},Y^{t}(W),J(W),S(W)}(z^{t-1},y^t_w,j(w),s(w)) \nonumber \\
&= \sum_{\substack{ y^n_{w,t+1} \in \mathcal{Y}^{n-t}}}\Big\{P_{Y^n_W}(y^n_w)\cdot P_{J(W),S(W)|Y^n_W}(j(w),s(w)|y^n_w) \nonumber \\
&~~~~~\cdot P_{Z^{t-1}|Y^n_W,J(W),S(W)}(z^{t-1}|y^n_w,j(w),s(w)) \Big\} \nonumber \\
&\overset{\mathrm{(d)}}= \sum_{\substack{ y^n_{w,t+1} \in \mathcal{Y}^{n-t}}}
\Big\{P_{Y^n_W}(y^n_w)\cdot P_{J(W),S(W)|Y^n_W}(j(w),s(w)|y^n_w)\nonumber \\
&~~~~~\cdot P_{Z^{t-1}|Y^n_W}(z^{t-1}|y^n_w) \Big\} \nonumber \\
&= \sum_{\substack{ y^n_{w,t+1} \in \mathcal{Y}^{n-t}}}\Big\{P_{Y^n_W}(y^n_w)\cdot P_{J(W),S(W)|Y^n_W}(j(w),s(w)|y^n_w)\Big\}\nonumber \\
&~~~~~\cdot P_{Z^{t-1}|Y^{t-1}(W)}(z^{t-1}|y^{t-1}_w) \nonumber \\
&= P_{Y^t(W),J(W),S(W)}(y^t_w,j(w),s(w))\nonumber \\
&~~~~~\cdot P_{Z^{t-1}|Y^{t-1}(W)}(z^{t-1}|y^{t-1}_w)
\end{align*}
\begin{align}
&\overset{\mathrm{(e)}}= P_{Y^{t-1}(W),J(W),S(W)}(y^{t-1}_w,j(w),s(w))\nonumber \\
&~~~~~\cdot P_{Y_t(W)|Y^{t-1}(W),J(W),S(W)}(y_{wt}|y^{t-1}_w,j(w),s(w)) \nonumber \\
&~~~~~\cdot P_{Z^{t-1}|Y^{t-1}(W),J(W),S(W)}(z^{t-1}|y^{t-1}_w,j(w),s(w)),
\end{align}
where
\begin{enumerate}[label=(\alph*)]
\setcounter{enumi}{3}
\item holds because $(J(W),S(W))$ is a function of $Y^n_W$,
\item follows because {of} the Markov chain
$Z^{t-1}-Y^{t-1}(W)-(J(W),S(W))$.
\end{enumerate}
Similarly, equation (\ref{zxxt}) can be shown as follows:
\begin{align}
&P_{Z^{t-1},X^{t}(W),J(W),S(W)}{(z^{t-1},x^t_w,j(w),s(w))} \nonumber \\
&= \sum_{\substack{ y^n_w \in \mathcal{Y}^n }}\Big\{P_{Y^n_W}(y^n_w)\cdot P_{J(W),S(W)|Y^n_W}(j(w),s(w)|y^n_w)\nonumber \\
&~~~~~\cdot P_{X^t(W)|Y^n_W,J(W),S(W)}(x^t_{w}|y^n_w,j(w),s(w))\nonumber \\
&~~~~~\cdot P_{Z^{t-1}|X^t(W),Y^n_W,J(W),S(W)}(z^{t-1}|x^t_{w},y^n_w,j(w),s(w)) \Big\} \nonumber \\
&\overset{\mathrm{(f)}}= \sum_{\substack{ y^n_w \in \mathcal{Y}^n }} \Big\{P_{Y^n_W}(y^n_w)\cdot P_{J(W),S(W)|Y^n_W}(j(w),s(w)|y^n_w)\nonumber \\
&~~~~~\cdot P_{X^t(W)|Y^n_W,J(W),S(W)}(x^t_{w}|y^n_w,j(w),s(w))\nonumber \\
&~~~~~\cdot P_{Z^{t-1}|X^t(W),Y^n_W}(z^{t-1}|x^t_{w},y^n_w) \Big\} \nonumber \\
&\overset{\mathrm{(g)}}= \sum_{\substack{ y^n_w \in \mathcal{Y}^n }} \Big\{P_{Y^n_W}(y^n_w)\cdot P_{J(W),S(W)|Y^n_W}(j(w),s(w)|y^n_w)\nonumber \\
&~~~~~\cdot P_{X^t(W)|Y^n_W,J(W),S(W)}(x^t_{w}|y^n_w,j(w),s(w))\Big\}\nonumber \\
&~~~~~\cdot P_{Z^{t-1}|X^{t-1}(W)}(z^{t-1}|x^{t-1}_{w}) \nonumber \\
&= P_{X^t(W),J(W),S(W)}(x^{t}_{w},j(w),s(w))\nonumber \\
&~~~~~\cdot P_{Z^{t-1}|X^{t-1}(W)}(z^{t-1}|x^{t-1}_{w}) \nonumber \\
&\overset{\mathrm{(h)}}= P_{X^{t-1}(W),J(W),S(W)}(x^{t-1}_w,j(w),s(w))\nonumber \\
&~~~~~\cdot P_{X_t(W)|X^{t-1}(W),J(W),S(W)}(x_{wt}|x^{t-1}_w,j(w),s(w)) \nonumber \\
&~~~~~\cdot P_{Z^{t-1}|X^{t-1}(W),J(W),S(W)}({z^{t-1}}|x^{t-1}_w,j(w),s(w)),
\end{align}
where
\begin{enumerate}[label=(\alph*)]
\setcounter{enumi}{5}
\item holds because $(J(W),S(W))$ is a function of $Y^n_W$,
\item follows due to {the} i.i.d.\ property of
each symbol and the Markov chain $Z^{t-1}-X^{t-1}(W)-Y^{t-1}(W)$,
\item follows because {of} the Markov chain
$Z^{t-1}-X^{t-1}(W)-(J(W),S(W))$.
\end{enumerate}
\qed

\section{\em Proof of Lemma \ref{leema}}

We will prove only (\ref{ztvt}) {by the well-known argument (cf.\ \cite{cover})}.
We introduce a timesharing variable
$Q$ which is uniformly distributed over $\{1,2,\cdots,n\}$ and is independent of all other RVs.
{The left-hand side of (\ref{ztvt})} can be rewritten as
\begin{align}
\sum_{\substack{ t=1 }}^nI(Z_{t};V_t)
&=n\left\{\frac{1}{n}\sum_{\substack{ t=1 }}^nI(Z_{t};V_t|Q=t)\right\} \nonumber \\
&= {n}I(Z_{Q};V_Q|Q) \nonumber \\
&= n[I(Z_{Q};V_Q,Q) - I(Z_{Q};Q)]\nonumber \\
&= nI(Z_{Q};V_Q,Q). \label{yquq}
\end{align}
By denoting $V=(V_Q,Q)$ {and} $Z = Z_{Q}$, (\ref{ztvt}) obviously holds.
The proof of {(\ref{ztut})--(\ref{xtut})} can be done similarly {by setting $X = X_{Q}$ and $Y = Y_{Q}$}.

To complete {the proof}, we need to verify that $Z_t-X_t(W)-Y_t(W)-U_t-V_t$ holds.
We shall first check that $Z_t-X_t(W)-Y_t(W)-{U_t}$ holds {for any $t \in [1,n]$}. To prove this claim,
we have to verify that
\begin{align}
&Z_t-X_t(W)-Y_t(W), \label{ztxtyt} \\
&X_t(W)-Y_t(W)-U_t, \label{xtytut} \\
&Z_t-(X_t(W),Y_t(W))-U_t. \label{ztxytut}
\end{align}
Indeed, Eqs.\ (\ref{ztxtyt}) and (\ref{xtytut}) clearly hold so the remaining task is to
check if the last one also holds.
Before checking that, we show that the Markov chain ${Z_t - (Z^{t-1},X_t(W),Y_t(W))-(J(W),S(W))}$,
which will be used {to confirm} (\ref{ztxytut}), holds.
\begin{align}
I(Z_t&;J(W),S(W)|{Z^{t-1}},X_t(W),Y_t(W)) \nonumber \\
&=H(Z_t|{Z^{t-1}},X_t(W),Y_t(W)) \nonumber \\
&~~~~~ - H(Z_t|{Z^{t-1}},X_t(W),Y_t(W),J(W),S(W)) \nonumber \\
&\overset{\mathrm{(i)}}\le H(Z_t|{Z^{t-1}},X_t(W),Y_t(W))  \nonumber \\
&~~~~~ - H(Z_t|{Z^{t-1}},X_t(W),Y^n_W,J(W),S(W)) \nonumber \\
&\overset{\mathrm{(j)}}= H(Z_t|{Z^{t-1}},X_t(W),Y_t(W))  \nonumber \\
&~~~~~ - H(Z_t|{Z^{t-1}},X_t(W),Y^n_W) \nonumber \\
&\overset{\mathrm{(k)}}= H(Z_t|X_t(W)) - H(Z_t|X_t(W)) \nonumber \\
&=0, \label{zero}
\end{align}
where
\begin{enumerate}[label=(\alph*)]
\setcounter{enumi}{8}
\item follows because conditioning reduces entropy,
\item holds because $(J(W),S(W))$ is a function of $Y^n_W$,
\item holds because each symbol of bio-data {sequences} is i.i.d.\
and we have $Z_t-X_t(W)-Y_t(W)$.
\end{enumerate}
From (\ref{zero}), it means that the conditional mutual
information is zero and thus ${Z_t - (Z^{t-1},X_t(W),Y_t(W))-(J(W),S(W))}$ forms a Markov chain.

Equation (\ref{ztxytut}) can be checked as follows:
\begin{align}
\hspace{-5mm}I(Z_t&;U_t|X_t(W),Y_t(W)) \nonumber \\
&= H(U_t|X_t(W),Y_t(W)) - H(U_t|X_t(W),Y_t(W),Z_t) \nonumber \\
&= H(Z^{t-1},J(W),S(W)|X_t(W),Y_t(W)) \nonumber \\
&\ \ \ \ \ - H(Z^{t-1},J(W),S(W)|X_t(W),Y_t(W),Z_t) \nonumber \\
&= H(Z^{t-1}|X_t(W),Y_t(W)) \nonumber \\
&\ \ \ \ \ + H(J(W),S(W)|X_t(W),Y_t(W),Z^{t-1}) \nonumber \\
&\ \ \ \ \ - H(Z^{t-1}|X_t(W),Y_t(W),Z_t) \nonumber \\
&\ \ \ \ \ - H(J(W),S(W)|X_t(W),Y_t(W),Z_t,Z^{t-1}) \label{48'} \\
&\overset{\mathrm{(l)}}= H(J(W),S(W)|X_t(W),Y_t(W),Z^{t-1}) \nonumber \\
&\ \ \ \ \ - H(J(W),S(W)|X_t(W),Y_t(W),Z^{t-1},Z_t)\nonumber \\
&\overset{\mathrm{(m)}}= H(J(W),S(W)|X_t(W),Y_t(W),Z^{t-1}) \nonumber \\
&\ \ \ \ \ - H(J(W),S(W)|X_t(W),Y_t(W),Z^{t-1}) \nonumber \\
&= 0,
\end{align}
where
\begin{enumerate}[label=(\alph*)]
\setcounter{enumi}{11}
\item holds because every symbol of bio-data sequences {is i.i.d.\ generated} so
the first and third terms in (\ref{48'}) {cancel} each other,
\item follows because ${Z_t - (Z^{t-1},X_t(W),Y_t(W))-(J(W),S(W))}$ holds (cf.\ (\ref{zero})).
\end{enumerate}
Thus, $Z_t-X_t(W)-Y_t(W)-U_t$ holds, and since $V_t$ is a function of $U_t$,
it follows that $Z_t-X_t(W)-Y_t(W)-U_t-V_t$ also forms a Markov chain.
\qed
\end{document}